\documentclass{ws-ijmpd}
\usepackage[super,compress]{cite}

\begin{document}

\markboth{V.~A.~Gani, A.~E.~Dmitriev, S.~G.~Rubin}
{Deformed compact extra space as dark matter candidate}

%
\catchline{}{}{}{}{}
%

\title{DEFORMED COMPACT EXTRA SPACE AS DARK MATTER CANDIDATE}
\author{VAKHID A.~GANI\footnote{Corresponding author, e-mail: vagani@mephi.ru}}
\address{Department of Mathematics, National Research Nuclear University MEPhI (Moscow Engineering Physics Institute), Kashirskoe shosse, 31, office A-211, 115409 Moscow, Russia}
\address{Laboratory of elementary particles theory, Theoretical Department, National Research Center Kurchatov Institute, Institute for Theoretical and Experimental Physics, 117218 Moscow, Russia}
\author{ALEXANDER E.~DMITRIEV\footnote{e-mail: alexdintras@mail.ru}}
\address{Department of Elementary Particle Physics, National Research Nuclear University MEPhI (Moscow Engineering Physics Institute), 115409 Moscow, Russia}
\author{SERGEY G.~RUBIN\footnote{e-mail: sergeirubin@list.ru}}
\address{Department of Elementary Particle Physics, National Research Nuclear University MEPhI (Moscow Engineering Physics Institute), Kashirskoe shosse, 31, office E-115, 115409 Moscow, Russia}

\vskip 1cm

\maketitle


\begin{abstract}
We elaborate the possibility for a deformed extra space to be considered as the dark matter candidate. To perform calculations a class of two-dimensional extra metrics was considered in the framework of the multidimensional gravity.
It was shown that there exists a family of stationary metrics of the extra space possessing point-like defect. Estimation of cross section of scattering of a particle of the ordinary matter on a spatial domain with deformed extra space is in agreement with the observational constraints.
\end{abstract}


\ccode{PACS Nos.: 04.50.-h, 95.35.+d}

\section{\label{sec:introduction} Introduction}

Last decades many attempts have been done to detect the dark matter \cite{history_UFN,dolgov_UFN,berez_UFN} as weakly interacting massive particles (WIMPs), but with no result up to now. Other dark matter candidates are topological defects which could bear energy (mass): domain walls, strings \cite{HKW01,KSS01} , vertices, dark atoms, etc. \cite{Khlopov01,Khlopov02}. The main purpose of our work is to consider space domains with the deformed small extra dimensions as dark matter particles.

Many problems of modern cosmology and the Standard Model as well as other fundamental questions can be clarified using the ideas of $f(R)$-gravity \cite{fR1,fR2} and extra-dimensional gravity \cite{randall02,dvali01,dienes01,dvali02,csaki01,csaki02,rubin01}, or the so-called ``vanishing dimensions'' scenarios \cite{stojkovic}. Strings, branes, multidimensional black holes and other additional entities open new perspectives in description of various physical phenomena. Nevertheless, many questions can be solved by using the extra space concept only.

Maximally symmetric metric of the extra space as a starting point are among the most popular in the literature. This assumption makes it possible to obtain clear and valuable results, see, e.g., \cite{2008IJMPD..17..785C,2011PhR...497...85M,2011PhRvD..84d4015B}. At the same time the space-time foam is supposed to be able to produce various geometries and there is no reason to assume that the extra-space geometry is simple \cite{1991PhLB..259...38L,Lindebook}. Random initial conditions --- the topology and metric of the manifold formed from the space-time foam --- play a key role in defining the properties of the extra space.

In this paper we study various stationary geometries of the two-dimensional sphere type. The Lagrangian we consider is a nonlinear function of the Ricci scalar that allows to stabilize the extra space size. The following situation is of specific interest for us: every point of a domain of the three-dimensional physical space has two compact extra dimensions with the extra space geometry differs from that of the ideal sphere. As we show, such a domain has an extra vacuum energy, i.e.\ it can be considered as a dark matter particle. We estimated the vacuum energy density as a function of the extra space deformation. We also discuss the cross section of such domains of small size interacting with particles of the ordinary matter.

Our paper is organized as follows. In Section~\ref{sec:extraction} we formulate the problem of gravitation in the six-dimensional space-time with two compact space dimensions. In Section~\ref{sec:metrics} we discuss the spherical topology of the extra dimensions: we formulate and solve the problem of the extra space configuration  dependence on the initial conditions. Possible phenomenology of the deformed extra space with the spherical topology is discussed in Section~\ref{sec:phenomenology}. We conclude with a discussion of our results and the prospects for future research.

\section{\label{sec:extraction} Separation of extra dimensions}

We will assume the characteristic size of the extra space to be small, and its geometry having quickly stabilized after the Universe was born. The stabilization is discussed in \cite{greene01,rubin02}.

Let's consider a Riemannian manifold
\begin{equation}\label{MD}
T\times M\times M'
\end{equation}
with the metric
\begin{equation}\label{metric}
ds^2 = g_{\mu\nu}(x)dx^{\mu}dx^{\nu} + G_{ab}(x,y)dy^a dy^b .
\end{equation} 
Here $T\times M$, $M'$ are manifolds with the metrics $g_{\mu\nu}(x)$ and $G_{ab}(x,y)$ respectively, and $T$ stands for the time dimension. Coordinates in the subspace $T\times M$ are denoted by $x$, while those in $M'$ by $y$. The indices $\mu$, $\nu$ ($a$, $b$) take values from 1 to 4 (5, 6, ...). The four-dimensional space-time $T\times M$ is called the ``main space'', while the $n$-dimensional compact space $M'$ --- the ``extra space''.

Hereinafter we consider a metric uniform with respect to the main space, so that $G_{ab}(x,y)=G_{ab}(t,y)$. The time dependence of the metric tensor $G_{ab}(t,y)$ is governed by the classical equations of motion, and can vary with initial or boundary conditions. At the same time, the energy dissipation in the main space $M$ makes the entropy of $M'$ decrease. This, in turn, leads to an effective friction term arising in the classical equations for the metric $G_{ab}(t,y)$, see \cite{rubin02} for more delails. This term stabilizes the extra space metric. On the other hand, as
\begin{equation}\label{stabi}
G_{ab}(t,y) \xrightarrow{t\rightarrow\infty}G_{ab}(y),
\end{equation}
the entropy of the extra space goes to its minimum, i.e.\ this process is accompanied by the decrease of the entropy and symmetrization of the extra space.

According to \eqref{metric}, the Ricci scalar is a sum of the scalar curvatures of the main 4-dimensional space-time and of the $n$-dimensional extra space:
\begin{equation}
R=R_4 + R_n
\end{equation}
with $G_{ab}$ independent of $x$.
Below we assume the natural inequality:
\begin{equation}\label{ll}
R_4 \ll R_n .
\end{equation}
Or more specifically:
\begin{equation}
R_4 =  \epsilon(x,y) R_n,\quad  \epsilon(x,y) \ll 1 \quad \text{for any }x,\: y.
\end{equation}
Let us consider the multidimensional gravity with higher derivatives and with the action given by (see, e.g., \cite{nojiri01,sokolowski01}):
\begin{equation}\label{act1}
S=\frac{m_D ^{D-2}}{2}\int d^4 x d^n y \sqrt{|G(y)g(x) |}f(R),\quad f(R) = \sum\limits_k {a_k R^k }
\end{equation}
with arbitrary parameters $a_k,\, k\neq 1$ and $a_1 = 1$. With the inequality \eqref{ll}, the function $f(R)$ in \eqref{act1} can be decomposed into the Taylor series:
\begin{equation}\label{FofR}
f(R)=f(R_n+R_4)\simeq f(R_n)+\epsilon(x,y) R_n f'(R_n).
\end{equation}
Then the action becomes
\begin{eqnarray}\label{act2}
&&S\simeq \frac{m_D ^{D-2}}{2}\int d^4 x d^n y\sqrt{|g(x)|} \sqrt{|G(y)|}[ \epsilon(x,y) R_n(x) f' (R_n (y) ) + f(R_n (y))] \nonumber \\
&& = \int d^4x  \sqrt{|g(x)|}\left[\frac{M^2 _{Pl}}{2}R_4 +  \frac{m_D ^{D-2}}{2}\int d^n y \sqrt{|G(y)|} f(R_n)\right],
\end{eqnarray}
where $D=n+4$. The Planck mass is given by the following combination
\begin{equation}\label{MPl}
M^2_{Pl}=m_{D}^{D-2}\int d^n y\sqrt{|G(y)|}f'(R_n(y)),
\end{equation}
and depends on the specific features of the asymptotically stationary geometry $G_{ab}(y)$.

According to the effective action \eqref{act2}, the cosmological $\Lambda$-term has the form:
\begin{equation}\label{density}
\Lambda =-\frac{m_D ^{D-2}}{2}\int d^n y \sqrt{|G(y)|} f(R_n).
\end{equation}
The parameters $a_k$ can be fine-tuned to make the $\Lambda$-term small enough in order to not spoil the agreement with the observations; we, however, will not address this issue in detail in this work.

The variation of the action with respect to the metric $G_{ab}(y)$ results in the following system of equations:
\begin{equation}\label{eqS1}
\frac{\delta S}{\delta G_{ab}(y)}= \frac{\delta S_1}{\delta G_{ab}(y)}+O(\varepsilon)=0, \quad S_1 = \frac{m_D ^2}{2}V_4 \int d^ny\sqrt{|G|}f(R_n),
\end{equation}
where $V_4 =\int d^4x\sqrt{|g(x)|}$. Neglecting terms proportional to the small parameter $\epsilon(x,y)$, we finally get:
\begin{equation}\label{eq2}
f'(R)R_{ab}-\frac{1}{2}f(R)G_{ab}-\nabla_{a}\nabla_{b}f'+G_{ab}\square f'=0.
\end{equation}
As the extra space is two-dimensional, only one of the two equations in \eqref{eq2} remains independent, which substantially simplifies the analysis of \eqref{eq2}. This equation can easily be obtained by taking the trace of the system \eqref{eq2}:
\begin{equation}\label{eqtrace}
\square\:\frac{df}{dR}=f-R\:\frac{df}{dR}\ ,
\end{equation}
where
\begin{equation}\label{squareGeneral}
\square=\dfrac{1}{\sqrt{G}}\partial_a\sqrt{G}G^{ab}\partial_b\ , \quad a,b=5,6.
\end{equation}
We solve this equation numerically, and the properties of the numerical solutions are discussed below. Eq.~\eqref{eqtrace} that describes the dynamics of the two-dimensional manifold, is a trivial identity if $f(R)$ is linear in curvature, $f(R)\propto R$. This equation does not have a solution if a source appears in its right-hand side.

\section{\label{sec:metrics} The metric of the extra space}

In line with the main idea formulated above, let us consider a metric corresponding to the extra space being of sphere topology:
\begin{equation}\label{sphereMetrix}
ds^2 = r^2(\theta)(d\theta^2 + \sin^2\theta\:d\phi^2).
\end{equation}
The Ricci scalar is now expressed via $r(\theta)$:
\begin{equation}\label{Rofr}
R(\theta)=\frac{2}{r^4\sin\theta}(-r'r\cos\theta+r^2\sin\theta+(r')^2\sin\theta-rr''\sin\theta),
\end{equation}
where the prime denotes differentiation with respect to $\theta$. In this case the operator \eqref{squareGeneral} has the form:
\begin{equation}\label{squareParticular}
\square =\dfrac{1}{r^2\sin\theta}\dfrac{d}{d\theta}\left(\sin\theta\dfrac{d}{d\theta}\right).
\end{equation}

Eqs.~\eqref{eqtrace} and \eqref{Rofr} can be considered as a system of two second order ordinary differential equations (ODE) for the functions $r(\theta)$ and $R(\theta)$. Obviously this system could be reduced to one fourth order ODE for the function $r(\theta)$. The system is solved in the segment $0\le\theta\le\pi$ with the appropriate boundary conditions at $\theta=0$.

Most works that consider compact extra spaces are based on maximally symmetric spaces with a constant Ricci scalar, the latter being a solution of the algebraic equation
\begin{equation}\label{eqtrace0}
f-R\:\frac{df}{dR}=0,
\end{equation}
which is a concequence of \eqref{eqtrace} and $R=const$. The solution $R=R_*$ of this equation of course could be obtained from the main system \eqref{eqtrace}, \eqref{Rofr} with the boundary conditions $R(0)=R_*$, $r(0)=\sqrt{2/R_*}$, $r'(0)=r''(0)=0$. 

To find the specific geometries of the extra space, we have to choose a form of the function $f(R)$. The simplest nonlinear in curvature theory is defined by
\begin{equation}\label{fR}
f(R)=U_1(R-R_0)^2+U_2,
\end{equation}
where $U_1$, $U_2$, and $R_0$ are arbitrary initial parameters. One-loop quantum corrections in the multidimensional quantum gravity for different configurations of the extra dimensions are discussed in \cite{QC}. The only restriction to these values is due to quantum effects which are meaningful at the scale below $m_D^{-1}>M_{Pl}^{-1}$. Classical description is valid if the size of the compact extra space $l\sim 1/\sqrt{R_*} \gg m_D^{-1}$, or
\begin{equation}\label{QuantLim}
R_* \ll 1
\end{equation} 
if $m_D =1$. The average curvature $R_*$ of the extra space is the result of solution of \eqref{eq2}, \eqref{fR}.  Hence, initial parameters have to be chosen to satisfy inequality \eqref{QuantLim}. It was also checked that the inequality $R_0 \ll 1$ is the most important, and we take it into account in our calculations, see, e.g., Fig.~\ref{fig:region} caption.

\begin{figure}
\center
	\includegraphics[scale=0.5]{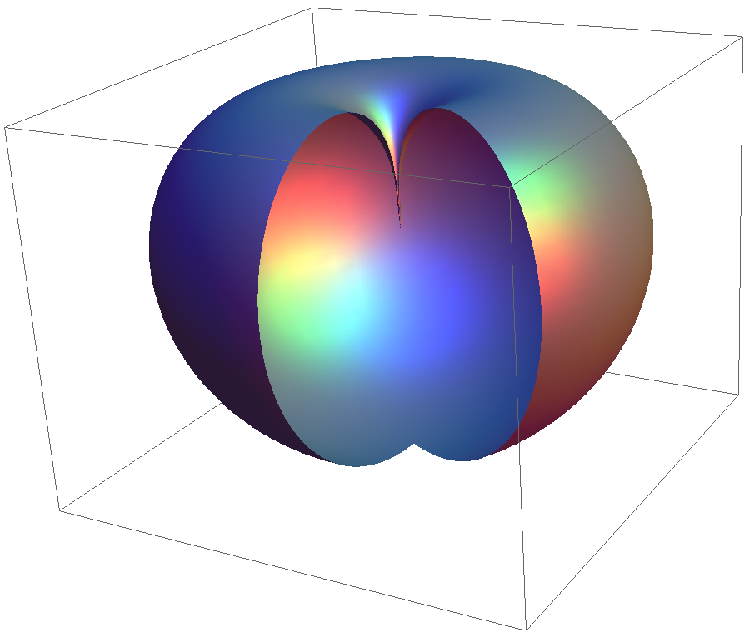}
	\hspace{1cm}
	\includegraphics[scale=0.5]{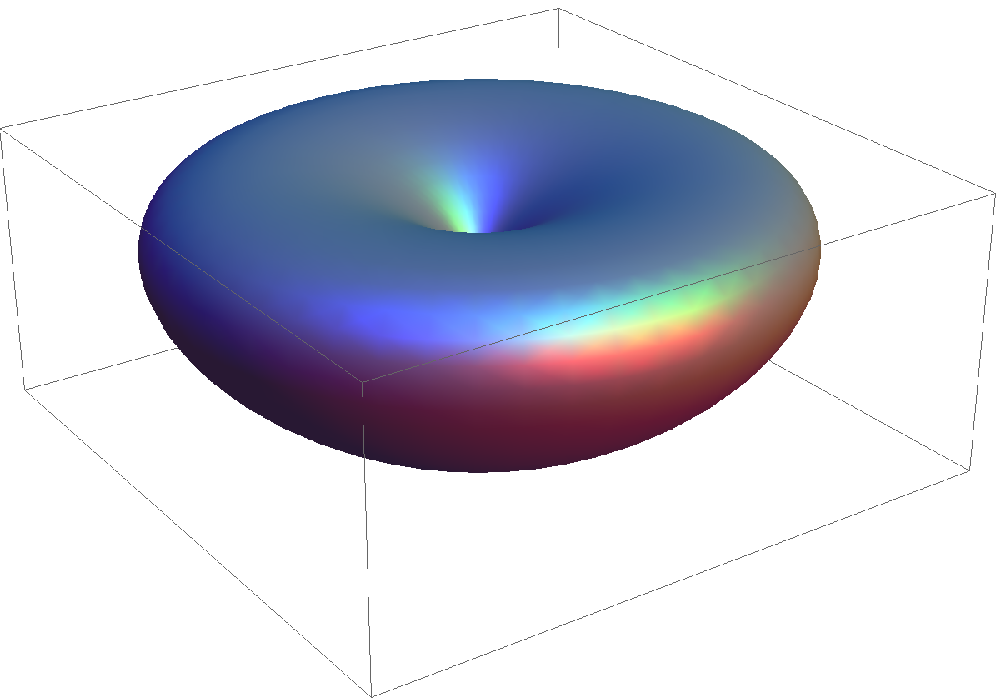}
	\caption{The extra space configurations of the ``apple'' type at $r''(0)=0.05$ (left panel) and $r''(0)=0.15$ (right panel).}
	\label{fig:fig1}
\end{figure}

\begin{figure}
\center
	\includegraphics[scale=0.5]{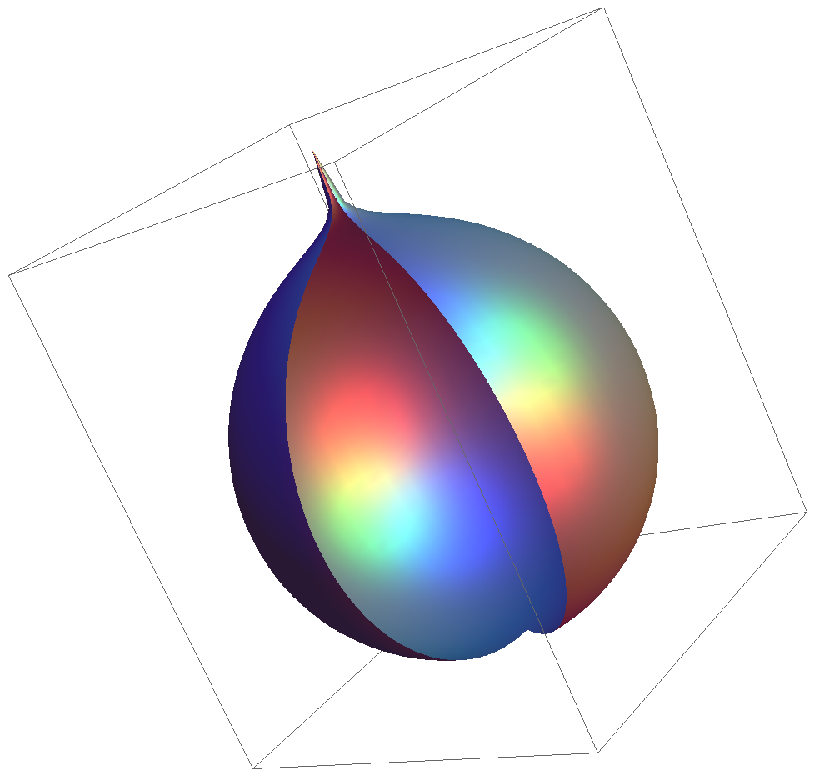}
	\hspace{1cm}
	\includegraphics[scale=0.5]{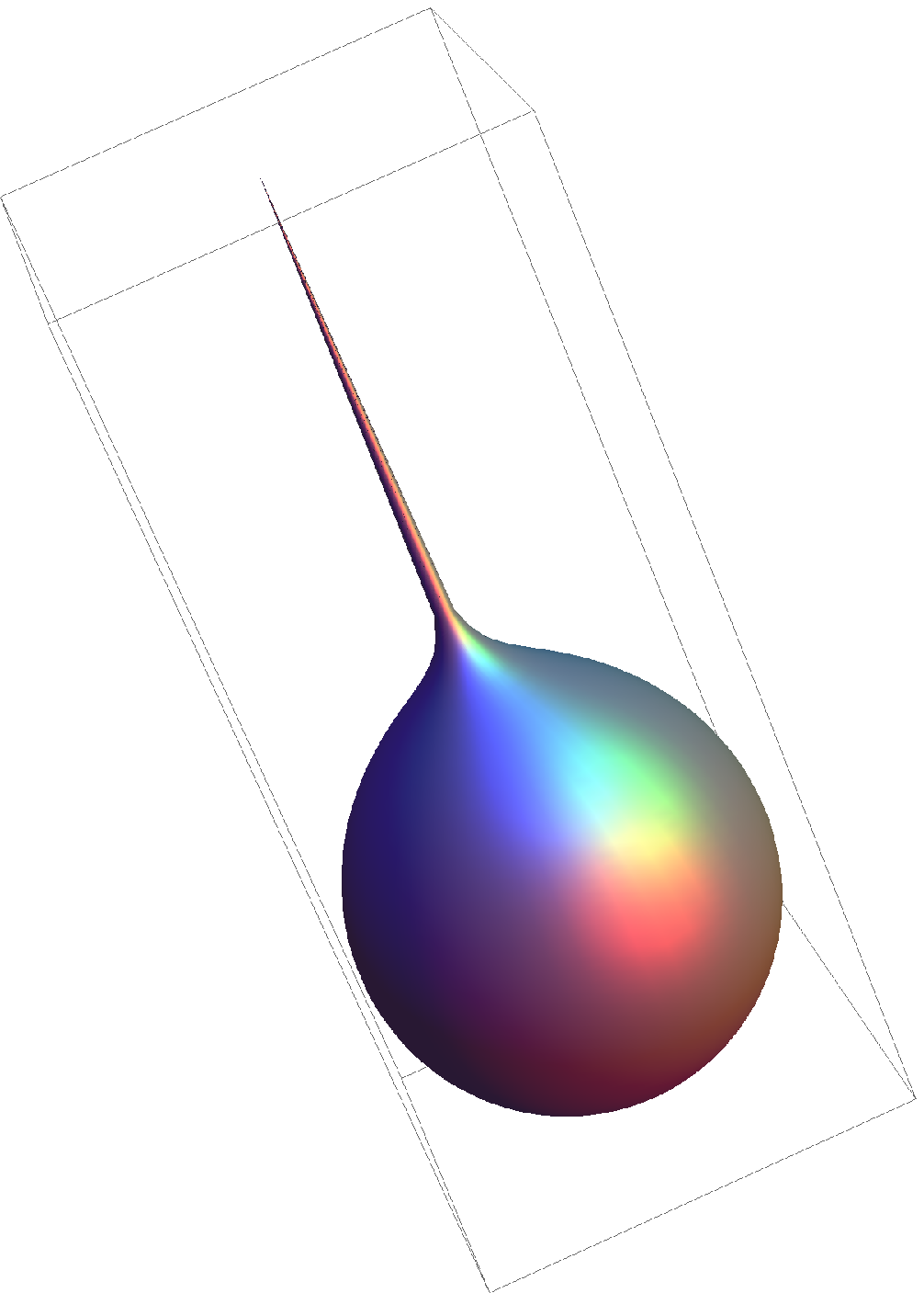}
	\caption{The extra space configurations of the ``onion'' type at $r''(0)=-0.01$ (left panel) and $r''(0)=-0.02$ (right panel).}
	\label{fig:fig2}
\end{figure}

Several characteristic numerical solutions of Eqs.~\eqref{eqtrace}, \eqref{Rofr} are presented in Figs.~\ref{fig:fig1} and \ref{fig:fig2}. Depending on the value of $r''(0)$ we get configurations of the ``apple'' type at $r''(0)>0$ and of the ``onion'' type at $r''(0)<0$. Apple-shaped geometries were discussed in \cite{gogberashvili}. The figures indicate the presence of a singularity at $\theta\to\pi$. Therefore, it is interesting to investigate the asymptotic behavior of the solution at $\theta\to\pi$ (or at $\theta\to 0$, due to the symmetry of the problem).

Let us look for the asymptotic solution in the form
\begin{equation}\label{rtet}
r(\theta)=\frac{C}{\theta^b}, \quad b>0,\quad\theta \rightarrow 0.
\end{equation}
Inserting it in \eqref{Rofr} we get:
\begin{equation}
R(\theta)=\frac{2(1-b/3)}{C^2}\theta^{2b}.
\end{equation}
As a result, Eq.~\eqref{eqtrace} takes the simple form
\begin{equation}
\frac{16U_1 b^2(1-b/3)}{C^4}\theta^{4b-2}=U_2+U_1 R_0^2,
\end{equation}
whence the parameters $b$ and $C$ may be found:
\begin{equation}\label{parbc}
b=\frac{1}{2}, \quad C=\left(\frac{3}{10}\left(\frac{U_2}{U_1}+R_0^2\right)\right)^{-1/4}.
\end{equation}
Thus a class of metrics with different boundary conditions at $\theta =\pi$ and the same asymptotics at $\theta=0$ is described by
\begin{equation}\label{asymp}
r(\theta)=\frac{C}{\sqrt{\theta}}, \quad \theta \rightarrow 0.
\end{equation}
The Ricci scalar
\begin{equation}\label{Ricci3}
R(\theta )=\frac{5}{3C^2}\:\theta , \quad \theta \rightarrow 0
\end{equation}
approaches zero in the vicinity of the point-like defect, hence the conical singularity is present at that point. This is because of $\delta$-functional source of matter in the Einstein gravitation, see \cite{vilenkin,kogan,navarro}.

\begin{figure}
\includegraphics[scale=0.5]{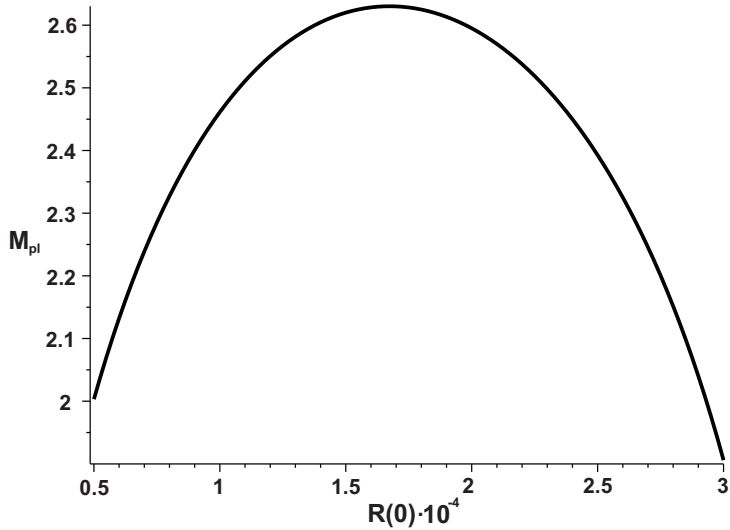}
\caption{The Plank mass dependence on the parameter $R(0)$, $U_2>0$. Here $M_{Pl}$ and $R(0)$ are in units of $m_D$ and $m_D^2$, respectively.}
\label{fig:mpl}
\end{figure}

\begin{figure}
\includegraphics[scale=0.5]{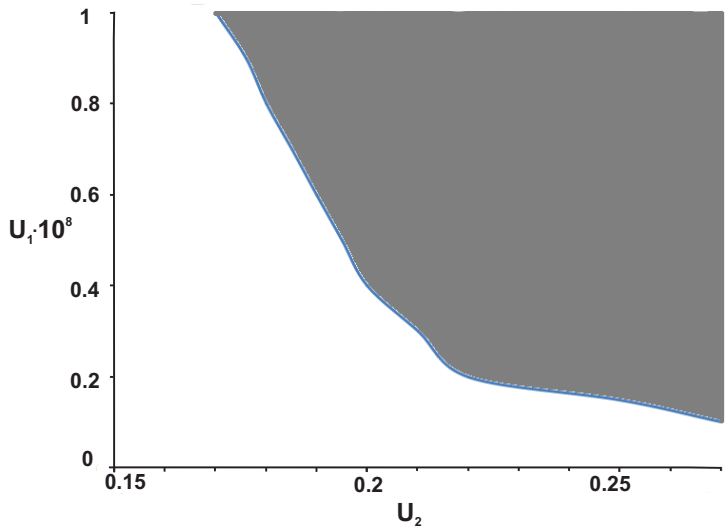}
\caption{The region of $U_1$ and $U_2$ where the Planck mass takes the observed value (here we used $R_0=\sqrt{3}\cdot 10^{-4}$); $U_1$ and $U_2$ are in units of $m_D^{-2}$ and $m_D^2$, respectively.}
\label{fig:region}
\end{figure}

The obvious fact that a set of boundary conditions (and hence a set of solutions of the equation) has the cardinality of the continuum leads to an interesting consequence: a set of the observable Planck masses \eqref{MPl} that depend on the extra space metric also has the cardinality of the continuum. So we arrive to a realization of the landscape idea, see, e.g., \cite{bousso}. The Planck mass dependence on the initial conditions is shown in Fig.~\ref{fig:mpl}. We get the observed value at $U_1=10^8\: m_D^{-2}$, $U_2=1\cdot m_D^2$ with the six-dimensional Planck mass of order of $m_D=10^{15}$ GeV of the GUT scale.

It is easy to show that there is a set of the parameters $U_1$ and $U_2$ such that the Planck mass has its observed value at some boundary conditions. The region of such values of $U_1$ and $U_2$ is shown in Fig.~\ref{fig:region}. Thus we come to the idea of the inverse landscape proposed in \cite{zinger,rubin03}, which essentially states that the observed values of the physical parameters do not unambiguously determine parameters of a primary theory, even if the primary parameters widely vary.

\section{\label{sec:phenomenology} Domains with the deformed extra space as particles of dark matter}

It is known that the extra space geometry effects could be interpreted as weakly interacting particles of dark matter \cite{panico01,kahil01}. Here we briefly discuss this issue using our approach. As shown above, properties of the extra space depend on the boundary/initial conditions. Within the model with the action \eqref{act1}, \eqref{FofR}, the maximally symmetric extra space with the metric \eqref{sphereMetrix} and hence $r(\theta)\equiv const$ can be formed as well as more complicated geometries with the metric \eqref{sphereMetrix}. The set of such geometries has the cardinality of the continuum.

\begin{figure}
\includegraphics[scale=0.5]{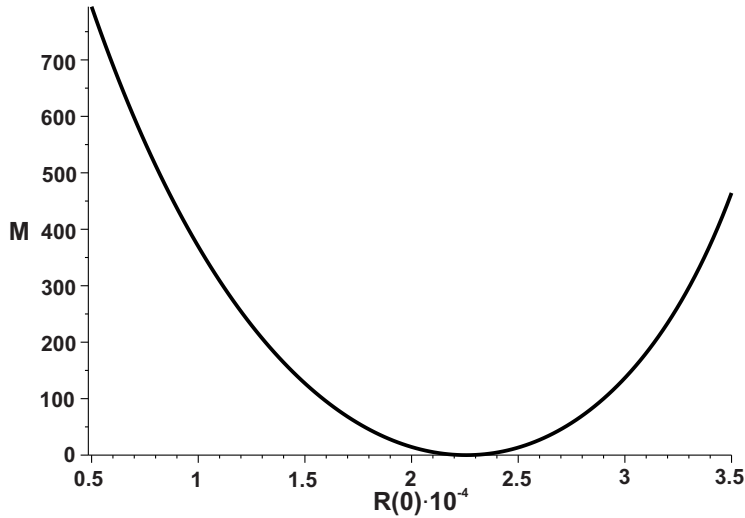}
\caption{The mass of the point-like defect as a function of the parameter $R(0)$, $U_2>0$. Here $M$ and $R(0)$ are in units of $m_D$ and $m_D^2$, respectively.}
\label{fig:mass}
\end{figure}

Suppose that the prior evolution resulted in the formation of a three-dimensional spatial domain $U$ of the main space with the extra space being of the ``apple'' or ``onion'' type at each point of $U$, while each point of the main space outside $U$ has the extra dimensions compactified into the ideal spheres, i.e.\ $r(\theta)=r_*=const$, $R=R_*=2/r_*^2$. Taking into account \eqref{FofR}, the extra space contribution to the energy density outside the domain becomes
\begin{equation}
\rho_0=m_D^4\pi\int r_*^2\sin\theta f(R_*)d\theta,
\end{equation}
while the corresponding contribution inside the domain is
\begin{equation}\label{energyDensityInside}
\rho=m_D^4\pi\int r^2(\theta)\sin\theta f(R(\theta))d\theta.
\end{equation}
Therefore a spatial domain of the main space with the extra space compactified into the ``apple''- or ``onion''-type manifolds will be observed as a domain with the extra energy density given above. Mass $M$ of such domain of a size $L$ can be estimated as
\begin{equation}
M\simeq (\rho-\rho_0)L^3
\end{equation}
for a distant observer. It is natural to suppose the minimal size of such domain be of the order of the extra space size $l$, i.e.\ less than $10^{-18}$ cm. In this case the mass of this particle-like object varies in wide range from zero to TeV scale and larger. Its dependence on the boundary conditions is shown in Fig.~\ref{fig:mass}.

The main feature of the dark matter is weakness of its interaction with particles of the ordinary matter. Let us estimate the strength of this interaction. As above, we assume the size of the domain to be of the order of the size of the extra space, $L\sim l\sim 10^{-18}$ cm. Consider the process of scattering of a particle of the ordinary matter on such point-like defect. In this paper we perform an estimation of the cross section of this process, working within the non-relativistic quantum mechanics.

The curvature is non-zero in a small vicinity of the point-like defect. Suppose the effective potential the particles are scattered on is
\begin{equation}
V(x)=V_0\exp{(-x^2 /l^2)}.
\end{equation}
In the Born approximation \cite{LL} the scattering cross section of a particle of mass $m$ is
\begin{equation}
\sigma \simeq \pi^2 m^2 V_0^2 l^6.
\end{equation}
As the only dimensional parameter is the extra space size $l$, we suppose
$$
V_0= Cl^{-1},\quad C\sim 1
$$
for estimation. The extra space size $l\lesssim 10^{-18}$ cm, therefore the scattering cross section for a particle of the mass of $1$ GeV is
\begin{equation}
\sigma \sim C^2\pi^2 m^2 l^4 \lesssim 10^{-43}\ \mbox{cm}^2.
\end{equation}
This estimate is consistent with the observational constraints on the mass and the interaction cross section of the dark matter particles.

\section{Conclusion}

In this paper we have considered stationary geometries of two-dimensional manifolds with the sphere topology. By itself, the possibility of stabilization of geometry in the absence of matter fields is a consequence of nonlinearity of the initial Lagrangian. As an example, we discussed the Lagrangian of the second power with respect to the Ricci scalar. We have shown that the sphere metric is a particular case of a more general class of metrics that are parameterized by the boundary condition at the regular center $\theta=0$. The numerical solution of the corresponding Cauchy problem showed that a singular point (topological defect) occurs at $\theta=\pi$.

The class of metrics found by us consists of two subclasses with different behavior at the singular point. The first class --- geometry of the ``onion'' type --- has rather simple asymptotic at the singular point that we have been able to analyse analytically. We showed that the Ricci scalar tends to zero with $\theta\to\pi$. At the same time the asymptotic form of the metric does not depend on the boundary conditions. The second class --- geometry of the ``apple'' type --- has more complicated structure and strongly depends on the boundary conditions. If this manifold is considered as the extra space extension of our four-dimensional space-time, new opportunities of low energy physics description appear. For example, the dark matter density becomes dependent on the extra space geometry.

Domain with the deformed extra dimensions has an extra energy density. Such domains interact with the ordinary matter by gravitation only, therefore they can be considered as dark matter candidate.

These dark matter domains move with the average velocity about $250$ km/s in the Galaxy. It is an interesting problem to find a method of their detection because ordinary particle detectors are not aimed for such events. Indeed, one can neglect the excitations of the domain during its non-relativistic interaction with the nucleus of the detector provided the multidimensional Planck mass is quite large. In this case pure elastic scattering of nuclei on the domain as a whole takes place. The question is whether one can distinguish between such a ``M{\"o}{\ss}bauer'' scattering and the ordinary one.

The minimal size of such objects is comparable to the size of the extra space with the mass varying in a wide range depending on the boundary conditions and the numerical values of the model parameters. A more or less natural mass of WIMPs in this model is about 10 TeV. In this case new methods to detect the dark matter have to be developed. The primary estimate of the interaction of such objects with the nucleons leads to the cross section about $10^{-43}$ cm${^2}$, which does not contradict the observational constraints for very massive dark matter particles.

In this paper we estimated the mass of the dark matter particle. To obtain its precise value, it is necessary to find a stationary solution of the Einstein equation in $4+2$ dimensions. We are going to study this question in the future.

One more problem discussed in this paper concerns the observable physical parameters, that are usually supposed to be uniquely connected with the initial theory parameters. Within multidimensional theories these are, e.g., the multidimensional Planck mass and the cosmological constant. But the situation is more interesting. We have shown that the observable Planck mass can be obtained depending on the initial (boundary) conditions at very different values of the initial theory parameters. The set of such primary parameters has the cardinality of the continuum.

\section*{ACKNOWLEDGMENTS}

This work was supported by the Ministry of Education and Science of the Russian Federation, Project No.~3.472.2014/K. The work of V.~A.~Gani was also supported by the Russian Federation Government under Grant No.~NSh-3830.2014.2. The authors are grateful to K.~A.~Bronnikov and V.~I.~Dokuchaev for useful discussions, and also to A.~I.~Bolozdynya and M.~D.~Skorokhvatov for their interest to this work.

\bibliographystyle{unsrt}

\end{document}